\documentclass[aps,prl,twocolumn,showpacs,superscriptaddress,floatfix]{revtex4}
\usepackage{graphicx}
\usepackage{dcolumn}
\usepackage{bm}
\usepackage{amssymb}
\usepackage{amsmath}
\newcommand{\req}[1]{Eq.~(\ref{#1})}
\newcommand{\reqs}[1]{Eqs.~(\ref{#1})}
\newcommand{\rref}[1]{(\ref{#1})}
\newcommand{\be}{\begin{equation}}
\newcommand{\ee}{\end{equation}}

\begin{document}

\title{Random resistor network model of minimal conductivity in graphene}
\author{ Vadim V. Cheianov}
\affiliation{Physics Department, Lancaster University,
  Lancaster, LA1 4YB, UK}
\author{Vladimir I. Fal'ko}
\affiliation{Physics Department, Lancaster University,
  Lancaster, LA1 4YB, UK}
 \author{Boris L. Altshuler} 
\affiliation{Physics Department, Columbia University, 538 West 120th Street, New
York, NY 10027, USA}
\affiliation{NEC-Laboratories America, Inc., 4 Independence Way, Princeton, NJ 085540, USA}
 \author{Igor L. Aleiner}
\affiliation{Physics Department, Columbia University, 538 West 120th Street, New
York, NY 10027, USA}
\date{June 19, 2007}

\begin{abstract}

Transport in undoped graphene is related to percolating current patterns in the 
networks of {\em N-} and 
{\em P}-type regions reflecting the strong bipolar charge density fluctuations. 
Transmissions of the {\em P-N} junctions, though small, are vital in establishing
the macroscopic conductivity.
We propose a random
resistor network model to  analyze scaling
dependencies of the conductance on the doping and disorder, 
the quantum magnetoresistance and the corresponding dephasing rate.
\end{abstract}

\pacs{73.23.-b, 72.15.Rn, 73.43.Qt, 81.05.Uw}
\maketitle

Graphene -- an atomic monolayer of graphite \cite{Geim} - is a gapless
semiconductor with linear \cite{Wallace} electron spectrum.
The carrier density 
in a graphene-based field effect transistor (GraFET)
 can be varied
continuously, from \textit{P}-type to \textit{N}-type.  
Experimental results \cite{Geim} are surprising:
the resistance
per square never exceeds several $k\Omega$'s 
in  contrast to the pinch-off in conventional (gapfull) semiconductors.
Such a behaviour of GraFET persists
over a broad temperature range $10^{-2} K \lesssim T\lesssim 10^2K$.
The conductivity reaches its minima  when the gate-controlled carrier
density, $n_g=0$, -- ``neutrality point''.
Similar observations were reported for the bilayer graphene  \cite{bilayer}.

Attempts to understand the finite value of the minimal conductivity in graphene have so far
addressed the role of ``chirality'' of the Dirac-type quasiparticles 
and influence of various short-range defects on the quantum
transport \cite{QuantumTr} in a homogeneous graphene sheet.
However, there is an emerging evidence \cite{Yacoby,Fuhrer,GeimKats}
that charge density in GraFETs is ``mesoscopically'' inhomogeneous -
probably, reflecting the fluctuations of the charge trapped
in the underlying substrate, or on its surface. An inhomogeneity of charge
density in a ``charge-neutral'' graphene sheet, $n_{g}=0$ implies that it can
be viewed as a checkerboard of \textit{N-} and \textit{P-}type doped regions
separated by  weakly conducting \cite{CheianovFalko}
{\em P-N} junctions.
In this Letter we propose a random resistor network model for such a
system and use it to describe classical and quantum transport.

\begin{subequations}
\label{model}
{\em Model} of the random resistor network (RRN) is formulated on the square
lattice with the lattice constant $a$ and sites labelled
by integers $(i,j)$.
Sites  $(i,j)$, $(i^\prime,j^\prime)$ with $|i'-i| > 1$, $|j'-j| > 1$ are not connected directly.
Each pair of sites  with $|i'-i|\leq 1$, and $|j'-j|\leq 1$,
is connected  by a link with the conductance
${\mathbb G}_{(i,j)}^{(i^\prime,j^\prime)}\!\!={\mathbb G}_{(i^\prime,j^\prime)}^{(i,j)}$ 
\begin{align}
&{\mathbb G}_{(i,j)}^{(i+1,j+1)}
={g}\left[1+(-1)^{i+j}\eta_{i,j}\right]/2;
\label{gnn}
\\
&{\mathbb G}_{(i,j+1)}^{(i+1,j)}
={g}\left[1-(-1)^{i+j}\eta_{i,j}\right]/2;
\label{gpp}\\
&{\mathbb G}_{(i,j)}^{(i+1,j)}={\mathbb G}_{(i,j)}^{(i,j+1)}=\gamma
g, \quad \gamma \ll 1.
\label{gnp}
\end{align}
Here $\eta_{i,j}$ is a random variable, 
\be
\eta_{i,j}=\pm 1,\ \langle\eta_{i,j}\rangle=p,  
\ \langle\eta_{i,j}\eta_{k,l}\rangle=\delta_{ik}\delta_{jl},
\ee
where $\langle\dots\rangle$ means the ensemble averaging.

\begin{figure}[h]
\includegraphics
[width =0.35 \textwidth] {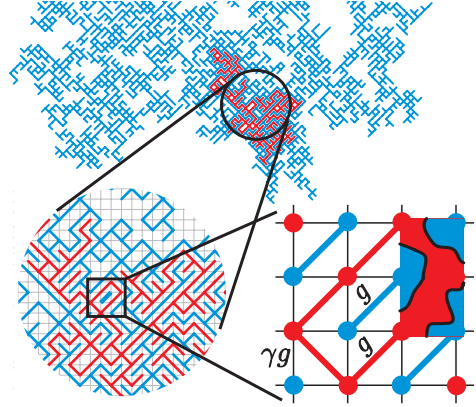}
\caption{Random resistor network (RRN) representation of a graphene sheet
with nominally zero doping, $n_{g}=0$.}
\label{fig1}
\end{figure}

The relation of model \rref{model} to 
graphene with inhomogeneous charge density, $n(r)=n_{g}+\delta n(\mathbf{r})$ is
illustrated on  Fig.~\ref{fig1}.
Assume that the density fluctuations  $\delta n(\mathbf{r})$ are characterised by the
length scale $a$  so that $\left\langle \delta n(\mathbf{r}
)\delta n(\mathbf{r+x})\right\rangle =\delta n^{2}f(x/a)$;
 $a$ may be determined by, \textit{e.g.},
the thickness of the insulating substrate.
As long as  $a^2\delta n\gg
1$, one can view the system as a combination of electron ({\em N}) and hole ({\em P}) puddles
of the size of the order of $a$. Each puddle contains a large number of carriers,
which are characterised by the Fermi wavevector  $k_{F}\sim \sqrt{\pi \delta n} \gg a^{-1}$.
If $i+j$ is even (odd), the site of 
 the RRN corresponds to  the {\em N}({\em P})-puddle and is marked by red (blue) colour on Fig.~\ref{fig1}.
The correspondence between the puddles and the lattice sites encodes the fact that the
 observable conductivity is determined
by the random links between the puddles rather than
the local conductivity of a puddle. 

 Each plaquette
of the RRN, see the inset for Fig.~\ref{fig1}, has one and only one diagonal connection --
either {\em P-P} or {\em N-N} link. This is described by \reqs{gnn} and \rref{gpp}.
If parameter $p=0$, then {\em P-P} and {\em N-N} connections appear
with equal probability, and $p>0\ (p<0)$ describe the electron (hole)
doping.
The boundary between the puddles has a finite, though
small, transparency \cite{CheianovFalko}. This transparency
is characterised by $\gamma \ll 1$ in \req{gnp} and depicted by the 
square-lattice grid on Fig.~\ref{fig1}.
\end{subequations}

{\em Scaling analysis} -- First, let us consider a RRN
with $p=0,\ \gamma=0$ in \req{model}, when 
two-colour random networks at a bipartite lattice have peculiar
geometrical features illustrated in 
Fig.~\ref{fig1} by a
computer-generated sample. 
Emerging patterns are typical for  the percolation theory \cite{PercolationReview}.
The RRN is
critical, {\em i.e.}  the geometry of the RRN is self-similar
on all length scales:
 $L\times L$  network contains,
typically, a larger cluster of one polarity (such as the red cluster in the
middle part of Fig.~\ref{fig1}) which separates a pair of smaller blue clusters. 
Each of those clusters is in turn a shell for several
smaller red clusters, etc.
Such alternating cluster-embedding represents a
scale-invariant property of an infinite network. As a result, larger and
larger parts of the network become excluded from the mono-colour percolation
upon the increase of the size $L$. Therefore, the observable conductance
$G(L)$ decreases with the increase of $L$. 
The corresponding critical behaviour of its mean value is \cite{PercolationReview}
\begin{equation}
\langle G(L) \rangle \sim (a/L)^{x}g; \quad  x\approx 0.97.
\label{G-P}
\end{equation}
This means that the conductivity is not defined.

Finite $p$ and $\gamma$ are relevant perturbations
for the percolation leading to a finite correlation length
$\xi(p,\gamma)$. At $L\gg \xi$, the RRN is
not critical 
and consists of 
independent patches of size $\xi$. Thus,  the 
  conductivity is finite:
\be
\sigma(\gamma,p) \equiv \langle G(L\to\infty) \rangle \sim \left[a/\xi (p,\gamma)\right]^xg.
\label{conductivity}
\ee

\begin{subequations}
Because of the scale invariance, $\xi$ depends on $p,\gamma$ as
\be
\xi\left(p,\gamma\to 0\right)\sim a |p|^{-\nu};
\ \xi\left(p\to 0,\gamma \right)\sim a \gamma^{-\mu}.
\ee
Its $p$-dependence is the dependence of the correlation length
on  deviations from the percolation threshold
\cite{PercolationReview}:
\be
\nu={4}/{3}.
\ee
 To the best of our knowledge,
the scaling of $\xi$ with $\gamma$ was not considered in
literature, yet. We found
\be
\mu={1}/{(h+x)}\approx 0.37,
\label{indices-b}
\ee
where $x$ is the coductance exponent, \req{G-P},
and $h=7/4$ (see Ref.~\cite{PercolationReview}) is the exponent of the
cluster
 outer perimeter  ${\cal P}(L)\simeq a (L/a)^h$  for
the cluster of the size $L$.
\label{indices}
\end{subequations}

To derive \req{indices-b},
 consider the red cluster of the size $L$ embedded into the conducting blue cluster.
Small but finite $\gamma$ does not affect the criticality
if the leak through the  perimeter of this
cluster $\mathcal{P}$ is much smaller than
the conductance of the blue cluster, the latter is also of the order of $G(L)$.
One can use \reqs{G-P}  and \rref{gnp} to write this condition as
$
(a/L)^x \gtrsim  (L/a)^h\gamma $.
This yields $L \lesssim \xi=a/\gamma^\mu$, with $\mu$ given by \req{indices-b}.
For $L > \xi$, the leakage through the 
cluster boundary  is efficient enough for the clusters to
become distinguishable, {\em i.e.} RRN  is uniform.

The dependence of the conductivity
on $p,\gamma\neq 0$, is
\be
\xi\left(p,\gamma\right)
\sim a \gamma^{-\mu}/ {\cal F}\left({p}/{p^*}\right),\quad
p^*=\gamma^{\mu/\nu},
\label{xi1}
\ee
where  ${\cal F}$ is a universal scaling function.
The form of \req{xi1} is protected by the scale invariance.
The
numerics  described below are well fit by the
interpolation formula
\be
{\cal F}(z)=\left(1+z^2\right)^{\nu/2},
\label{crap}
\ee
reproducing both $z\ll 1$ and $z\gg 1$ asymptotic behaviour. 
Substituting \reqs{xi1}, \rref{crap} into \req{conductivity}
we find
\be
\sigma(\gamma,p)= u g \gamma^\alpha {\cal F}^x
\left({p}/{p_*}\right), \quad
\alpha=x\mu\approx 0.36,
\label{finalsigma}
\ee
where $u\approx 1.3$ is a coefficient found from the
fit of the numerical data.
Equation \rref{finalsigma} completely describes non-analytic
dependence of the conductivity on both the leakage parameter $\gamma\ll 1$ \cite{For_you_Idiots}
and adjustable doping $p$. 
Recently, an attempt was made in Ref.~\cite{DasSarma} to explain the graphene minimal
conductivity using a mean field theory. We believe that Ref.~\cite{DasSarma} correctly
describes the limit of high carrier density, where the 
fluctuations of local conductivity are small, but fails near the neutrality point
where the percolation physics start to dominate.

\begin{figure}[ht]
\includegraphics
[width =0.33 \textwidth]
{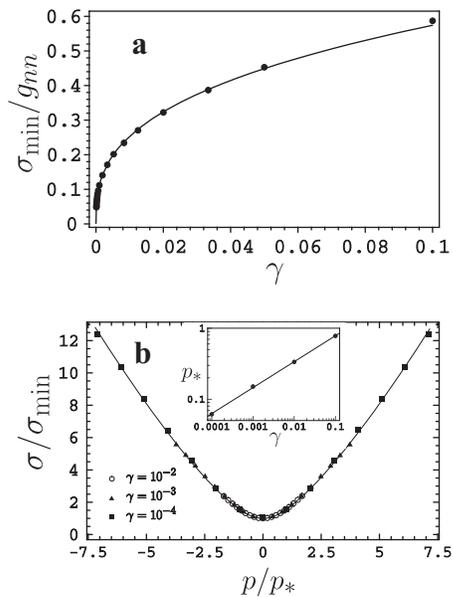}
\caption{(a) Conductivity $\sigma _{\min }(\gamma)\equiv\sigma(\gamma,p=0)$. 
 Numerical data are
represented by dots. The solid line is the best fit to the scaling law in
\req{finalsigma}. \ (b) Collapse of the conductivity data
obtained for RNNs with various $-1/2<p<1/2$  and values of the
parameter $\protect\gamma $ onto a single curve, \req{crap},
 represented by solid line. For each value of $\gamma$
the corresponding values of the parameter  $p_{\ast }$\ were found
using the by fitting $\protect\sigma (p)$ to \reqs{finalsigma}
and \rref{crap}, see inset.}
\label{fig2}
\end{figure}

{\em Numerical simulations} 
of model \rref{model}
were performed on square lattices with $L\leq 500a
$. Numerical evaluation of the network conductance has been done using the
bus-bar boundary conditions at the two opposite edges along the direction of
the current flow and the periodic boundary conditions in the orthogonal
direction. Realizations of the RRN were generated in
families spanning the interval $p\in \lbrack -1;1]$ of the network
parameter, in steps of $\Delta p=0.0125$. Each family was obtained from the
regular network at $p=1$ (containing only blue links) by sequentially
replacing $L^{2}/\Delta p$ blue links by red ones in randomly chosen
plaquettes. Averaging of the data has been performed over 300 families. 
To minimise the finite-size effects we 
(i) identified $1/L$ corrections by finite size scaling from $L=50a$ to $L=500a$; and
(ii) subtracted those corrections by means of numerical extrapolation. 
Numerical results  summarised on Fig.~\ref{fig2}
are in excellent agreement with the scaling formulas \rref{crap}, \rref{finalsigma}.

\begin{figure}[ht]
\includegraphics
[width =0.31 \textwidth]
{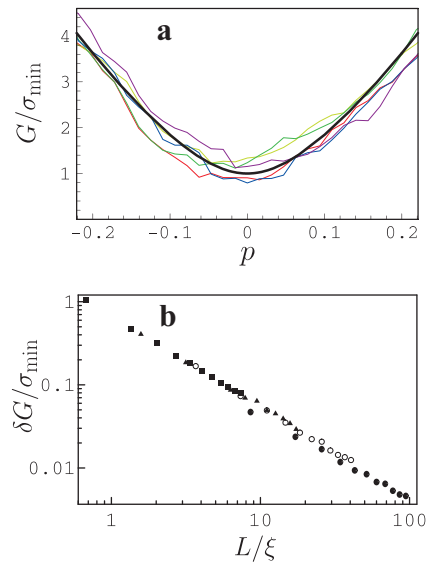}
\caption{(a) Conductance of several 100$\times $100 samples with $\protect%
\gamma =10^{-4}$ as a function of $p$: \ The parameter $p$ is gradually
changed from -.2 to 0.2 by flipping a small number of randomly chosen links
from \textit{P-P} type to the \textit{N-N } type. The fluctuations of the
conductance are the strongest at $p=0$. (b) R.m.s. of sample-to-sample
fluctuations of RRN conductance for $p=0$ and various values of $\protect%
\gamma $ as a function of system size, $L$. }
\label{fig3}
\end{figure}

{\em Sample-to-sample fluctuations} --
Geometry of the percolation cluster is non-trivial. In particular,
its linear size depends on the particular realization and 
re-conecting only few bottleneck links may change it substantially. Therefore, the conductance
$G(L)$ is a random quantity, which is characterised by its distribution function
$P(G)$ rather than only by its average. Scaling invariance of the critical cluster
at $L\ll \xi(p,\gamma)$ constraints the functional form of this distribution to
$
P(G)={\langle G\rangle}^{-1}{\mathbb P}\left({G}/{\langle G\rangle}\right),
\ \int {\mathbb P}(x)dx=1,
$ 
where ${\mathbb P}(x)$ is a universal function. 
It implies the variance
\be
\langle \delta G^2(L\ll\xi)\rangle =u_2\langle  G(L\ll \xi)\rangle^2, 
\ee
where $u_2\simeq 1$ is a universal numerical coefficient.
For $L\gg\xi$, the central limit theorem is
restored, and we find
\be
\langle \delta G^2(L\gg \xi)\rangle= \langle \delta G^2(\xi)\rangle \left({\xi}/{L}\right)^2=
u_2 \sigma^2 *  \left({\xi}/{L}\right)^2.
\label{variance}
\ee
Figure~\ref{fig3} (a) illustrates the 
 fluctuations of RRN conductance. One can see that
not only value of the minimal conductance but also its position fluctuates
from sample to sample. Figure~\ref{fig3} (b) shows an excellent agreement
with the scaling form \rref{variance} for $p=0$.

The fluctuations \rref{variance} are self-averaging. Notice, however, that
because $\xi$ is large, those fluctuations can exceed the
universal conductance fluctuations $\simeq e^2/\hbar$ \cite{UCF} which are not
self-averaging as long as the phase coherence is preserved.
The scale for the energy dependence of fluctuations \rref{variance} is
determined by the properties of the individual bottle-neck links in the cluster and
significantly exceeds the  Thouless energy for the UCF.
This observation is consistent with numerical findings of
Ref.~\cite{Beenakker}.

{\em Quantum magnetoresistance} --
Up to now we ignored the quantum interference and interaction effects.
Some, and most importantly those that give rise to the quantum magetoresistance \cite{WL},
can be analysed within the percolation cluster framework \cite{DKh}.
The role of the weak localisation (WL) is to renormalize
the conductance of each link in \req{model} due to the interference
of the random walk paths that go through the same link twice. For WL magnetoresistance only long
paths are important. Thus, one can assume that the return 
probability is independent of each particular links, so that it
can be included into the renormalization of the  conductance $g$.

As usual \cite{WL}, the evaluation of the WL correction consists in 
finding the classical return probability for the random walk, whereas
the quantum mechanics determines the relevant spatial scales and
the pre-factors in this problem, which are the same as in the case
of  homogeneous graphene \cite{WLMcCann}.

To find the probability for the returning path,
one poses the problem of the diffusion on the lattice, \req{model},
where the link conductances ${\mathbb G}$ are replaced by the diffusances.
 Then, comparison with \reqs{G-P}, \rref{conductivity}
yields the scale-dependent diffusion coefficient $D(L)$ in terms of
the observable diffusion constant at large distances, $D_\xi$:
\be
 D(L) \sim  [\xi/{\rm min}\,(L,\xi)]^{x} D_\xi.
\label{diffusion}
\ee

Our treatment of the WL differs from
that of Ref.~\cite{WLMcCann} only by the scale dependence of the diffusion constant.
The magnetoconductance $\delta \sigma(B)=\sigma(B)-\sigma(0)$ in not-so-low
magnetic field $B$
can be estimated as
\begin{align}
&\delta\sigma \approx \frac{e^2}{\pi^2\hbar}\sum_{J=0}^1(-1)^J
\!\!\!
\sum_{M=-J}^J
\!\int_0^{\sqrt{eB/\hbar c}}\!\!\!\!\frac{qdq\, D_\xi}{D(1/q)q^2+\Gamma_J^M},
\label{WL}
\\
&
{\Gamma}_0^0=\frac{1}{\tau_\phi};\
\, {\Gamma_1^M}=\frac{1}{\tau_\phi}
+\frac{1}{1+|M|}\left(\frac{1}{\tau_\perp}+\frac{|M|}{\tau_\parallel}\right).
\nonumber
\end{align}
Here $\tau_\perp$ is the inter-valley scattering time and $1/\tau_\parallel$ is the
rate for intra-valley scatterings breaking certain symmetries of the system,
see Ref.~\cite{WLMcCann} for more details.
The  phase relaxation time
$\tau_\phi$ will be estimated below.

If $\tau_\phi \gg \xi^2/D_\xi$, \req{WL} reduces to the results of Ref.~\cite{WLMcCann}.
The specifics of the scale dependent diffusion \rref{diffusion} 
are revealed in the opposite limit $\tau_\phi \ll \xi^2/D_\xi$:
\be
\begin{split}
&\delta\sigma \approx \frac{e^2}{\pi^2\hbar}\sum_{J=0}^1(-1)^J
\!\!\!
\sum_{M=-J}^J\!
\left(\frac{{\cal L}_{J,M}}{\xi}\right)^{x}
\tilde{Y}\left(\frac{eB{\cal L}_{J,M}^2}{c \hbar}\right);\\
& 
{\cal L}_{J,M}=\xi \left[{D_\xi}/{(\xi^2\Gamma_J^M)}\right]^{\frac{1}{2+x}},
\end{split}
\raisetag{8mm}
\label{WLsmall}
\ee
where $\tilde{Y}(z)$ is a universal function \cite{caution}
with the asymptotic behaviour $\tilde{Y}(z\ll 1) \simeq z^2$,
and $\tilde{Y}(z\gg 1) \approx 1-z^{-x/2}$.
The overall magnetoresistance \rref{WLsmall} is strongly suppressed
in comparison with that for the homogeneous graphene,
$|\delta \sigma|\ll e^2/(\pi^2\hbar)$.

The phase relaxation in ordinary disordered conductors is dominated by
the electron-electron interaction \cite{AAG}. The  time defining
the magnetoresistance curvature  $\partial^2 \sigma/\partial B^2|_{B \to 0}$
is controlled by the dimensionless conductance ${\cal G}(L)=2\pi\hbar G(L)/e^2$:
\be
{\hbar}/{\tau_\phi}\simeq {T}/{{\cal G}(L_\phi)}; \ L_\phi=\left(D\tau_{\phi}\right)^{1/2}, 
\label{tauphi}
\ee
where $L_\phi\equiv {\cal L}_{0,0}$ is the dephasing length. As
the all the geometric properties of the system in \req{tauphi} are encoded into
the scale dependences of ${\cal G}(L)$ and $D(L)$,  \req{tauphi}
is valid even for diffusion along the critical cluster. We obtain
from \reqs{tauphi}, \rref{G-P}, \rref{conductivity}, and \rref{diffusion}
\be
{L_\phi}= 
\xi \Phi\left({T_\xi/T}\right);
\ T_\xi={\hbar{\cal G}(\xi)D_\xi}/{\xi^2} \propto \xi^{-2(1+x)},
\label{Lphi}
\ee
where $\Phi(x)$ is a scaling function with the asymptotic
behaviour $\Phi(z\gg 1)\approx \sqrt{z}$, $\Phi(z\gg 1)\approx {z}^{1/(2+2x)}$.
Condition $T \sim T_\xi$ determines the crossover from the critical to
the normal diffusion: $L_\phi(T>T_\xi)>\xi,\ L_\phi(T<T_\xi) < \xi$.

Equations \rref{Lphi} and \rref{WLsmall} predict interesting
doping behaviour of the magnetoresistance at fixed temperature.
At $p \lesssim 1$, $\xi(p)$ is small, $T_\xi > T$ and the usual logarithmic magnetoresistance
occurs. For $p\to 0$, $T_\xi < T$, and the WL becomes suppressed as $1/\xi^x$.
Such prediction seems to be consistent with the experiment of Ref.~\cite{WLGeim}.


{\em Relation to the experimental parameters} --
To relate \req{finalsigma} to the
 properties of graphene, we have to connect
the parameters $\gamma$ and $g$ with the physical parameters of
the sample.
If
the charge inhomogeneity is the dominant disorder in the
graphene monolayer, we can
estimate the conductance of \textit{N-N} (\textit{P-P}) connections between
puddles as $g\sim \frac{e^{2}}{\hbar } ak_{F}$ and conductance of the {\em P-N}
junction separating puddles of the opposite polarity as $\gamma g\sim 
\frac{e^{2}}{\hbar}(ak_{F})^{1/2}$, see Eq.~(2) of
Ref.~\cite{CheianovFalko},
 $\gamma \sim
\left( ak_{F}\right) ^{-1/2}\sim \left( a^{2}\delta n\right) ^{-1/4}$.
According to \req{finalsigma} we then estimate $
\sigma _{\min }=\sigma(\gamma,p=0) \sim \frac{e^{2}}{\hbar}\left( a^{2}\delta n\right)^{\frac{1}{2}-%
\frac{\alpha }{4}} \sim \frac{e^{2}}{\hbar}\left( a^{2}\delta n\right)^{0.41}$.

{\em In conclusion}, we constructed a random resistor network model, which adequately
takes into account strong fluctuations of the local charge density and,
thus, of the local conductivity of the mono- and bilayer graphene near neutrality point.
 This model describes inhomogeneous
current percolating through the system, giving rise to the  scaling dependencies
of the observable conductivity on the doping and disorder. Quantum magnetoresistance
and the sample-to-sample fluctuations are  analysed within the model.

We are grateful to A.K. Geim and P. Kim for discussions.


\begin{thebibliography}{99}
\bibitem{caution}Estimate \rref{WL} reproduces $\tilde{Y}(z\gg 1)$. At small
magnetic field, $z\ll 1$, \req{WL} can not be used.
\end{thebibliography}
\end{document}